\documentclass[runningheads,a4paper]{llncs}

\usepackage{amssymb}
\usepackage{graphicx}
\usepackage{soul}
\usepackage{url}
\usepackage[natbibapa]{apacite}
\usepackage{tipa}
\usepackage{appendix}
\newcommand{\keywords}[1]{\par\addvspace\baselineskip
\noindent\keywordname\enspace\ignorespaces#1}

\usepackage{float}

\usepackage{xcolor}

\begin{document}

\mainmatter  % start of an individual contribution

\title{An Exploratory Study on Perceptual Spaces of the Singing Voice}

\author{Brendan O'Connor\and Simon Dixon\and George Fazekas \thanks{This research is funded by the EPSRC and AHRC Centre for Doctoral Training in Media and Arts Technology (EP/L01632X/1). }}

% if the names of the authors are too long for the running head, please use the format: AuthorA et al.
\authorrunning{Brendan O'Connor, Simon Dixon and George Fazekas}
\titlerunning{{\em 2020 Joint Conference on AI Music Creativity}, Full Paper}

\institute{Centre for Digital Music, Queen Mary University of London, UK\\ \email{\{b.d.oconnor, s.e.dixon, g.fazekas\}@qmul.ac.uk}}

\maketitle

\begin{abstract}

Sixty participants provided dissimilarity ratings between various singing techniques. Multidimensional scaling, class averaging and clustering techniques were used to analyse timbral spaces and how they change between different singers, genders and registers. Clustering analysis showed that ground-truth similarity and silhouette scores that were not significantly different between gender or register conditions, while similarity scores were positively correlated with participants' instrumental abilities and task comprehension. Participant feedback showed how a revised study design might mitigate noise in our data, leading to more detailed statistical results. Timbre maps and class distance analysis showed us which singing techniques remained similar to one another across gender and register conditions. This research provides insight into how the timbre space of singing changes under different conditions, highlights the subjectivity of perception between participants, and provides generalised timbre maps for regularisation in machine learning.

\end{abstract}

\section{Introduction}

% \pagenumbering{gobble}

The human voice is arguably one of the most diverse instruments available to musicians, making use of its own complex structure of source excitation, filtering and articulation functions to produce a wide variety of vocal sounds. In this paper we investigate how the perception of vocal timbre space changes under different listeners biases and singer conditions. We challenge the standard taxonomy of vocal techniques, as labelled in the VocalSet dataset \citep{wilkins2018}, and consider how a listener's musical background affects this. We use psuedo-randomly chosen samples from 6 of this dataset's singers to represent 5 singing techniques under the different conditions of singer identity, gender and register. Participants produced pairwise dissimilarity ratings between these vocalisations, and the results were analysed using multidimensional scaling, class averaging and cluster analysis. The source code, stimuli and collected data are available online.\footnote{https://github.com/Trebolium/VoicePerception}

\section{Related Work}

\subsection{Voice Production} \label{secVoicePhysio}

There is much literature describing how the mechanics of the voice organs combined with individual morphological differences affect vocal sounds \citep{garcia-lopez2010, sundberg1987, kayes2015, zhang2016}, providing detailed insight into how vocal production techniques influence perception of a singer's voice. A survey of voice transformation techniques by \citet{stylianou2009} discusses interdependence between vocal mechanisms and how it is a vital consideration when building a model of the voice.
\citet{garcia-lopez2010} compare values and perspectives between artistic and scientific professions specialising in the voice and observe that differences between these two communities lead to a convoluted and inconsistent tapestry of technical terminology often leading to mislabelling or misunderstanding vocal production processes - an observation shared by many others \citep{sundberg1987, gerratt2001, sundberg1981, proutskova2019}. The term `phonation modes' classifies specific configurations of the voice organs that lead to a particular timbre quality in the voice \citep{sundberg1987}. \citet{proutskova2013} assert that phonation modes are not linked to singing registers, introducing the question of how changes in pitch affect the timbre while this subset of vocal techniques remains constant.
\subsection{Perceptual Studies}

An intuitive method for building timbre maps of an instrument is to collect perceptual dissimilarity data between audio clips. These can be formatted into dissimilarity matrices and converted into a representation of fewer dimensions via Multidimensional Scaling (MDS). MDS is especially useful for representing the cognitive process of how humans perceive and generalise the diversity of data within a given domain \citep{mugavin2008}. The first to use MDS for perceptual representations were \citet{kruskal1964, shepard1962, shepard1962a}, employing `nonmetric' MDS techniques (due to the rank-ordered nature of the data) to reflect the data monotonically in the MDS representations, which has been commonly used for investigating timbre spaces \citep{gerratt2001, mcadams1995, wedin1972, krimphoff1994, serafini1993}.

MDS has been adapted for different uses over the years. \citet{carroll1970} improved on classical MDS with the algorithm INDSCAL (used by \citet{grey1977}) which avoids rotational invariance for simplified dimensional interpretation and provides weights relating the contribution of participants' collected data to these dimensions \citep{mugavin2008}. Interpreting the timbral meaning of MDS dimensions often requires a post-hoc analysis to find correlations to audio descriptors \citep{krimphoff1994}. \citet{mcadams1995} however, combined perceptual dissimiliarities with acoustic parameters to generate timbre maps using the CLASCAL algorithm \citep{winsberg1993}, which greatly improved dimensional interpretation.

\citet{iverson1993} investigated the influence of entire tones and their corresponding onsets/remainders on timbre spaces, concluding that the salience of acoustic attributes in entire tones cannot be attributed to either their onsets or remainders alone. Interestingly, it has been commonly reported that elements such as attack transients assisted in \textit{identifying} an instrument, yet without affecting perceptual structures between the instruments/classes \citep{wedin1972, mcadams1995,iverson1993,krimphoff1994, grey1977}.

%\subsubsection{Clustering}

\citet{mcadams1995} used nearest-neighbour clustering analysis to detect significant differences between participants, highlighting instances where some individuals may have misinterpreted instructions. \citet{gerratt2001} used the K-means algorithm to confirm that their choice of dimensionality separated their three classes into statistically significant clusters. \citet{grey1977} did similar calculations and applied the HICLUS hierarchical clustering algorithm \citep{johnson1967} to group the stimuli into clusters and assess the \textit{compactness} of these clusters. \citet{iverson1993} averaged dissimilarity values across all participants' perceptual data in order to generate MDS on averaged values, although it may have been beneficial to use INDSCAL's participants' weight values to confirm whether there were outlier participants in the data.

%\subsubsection{Participant Management}

\cite{grey1977} reports that the order in which comparisons are represented causes differences in judgements between participants, and it is therefore common practice to randomise the presented order of pairwise comparisons. \citet{mehrabi2018, gerratt2001} include repeated examples for rating within experiments to assure intra-participant reliability and consistency. In relation to how participant profiles affected their rating techniques, \citet{wedin1972, carterette1974} found that participants' different levels of musical training did not lead to major systematic differences between them. \citet{mcadams1995} noted similarly, but observed that more musical participants achieved more precise ratings. However \citet{serafini1993} reported that musicians familiar with the gamalan sounds being evaluated attributed more importance to the attack of the sound than its resonant volume, while non-musicians considered these aspects equally. 

\section{Experiment} \label{ExpChapter}

% different no classes, learn how general or specific a model must be
The literature referenced in Section \ref{secVoicePhysio} describes disagreement and confusion between professionals specialising in the voice when describing and ascribing vocal techniques. As a result of this, we hypothesise that participants' dissimilarity ratings will cause clusters to diverge significantly from those implied by the ground truth labels. Observing how much variance there is in vocal timbre space between the different conditions of gender, pitch, singer identity and participant musicality will allow us to assess how generalisable a single model of a singing voice can be.

\subsection{The Stimuli} \label{dataSampling}

% who sings what, how many examples
3 male and 3 female singers (identified in Section \ref{secResDisc}) were randomly selected from VocalSet \citep{wilkins2018}, a dataset containing audio of singers vocalising a range of pitches and sustained vowels, annotated by the different vocal techniques being used. The techniques \textit{straight, belt, breathy, fry, vibrato} were selected based on their frequent occurrence in popular Western singing. We extracted 2 separate sets of vocalisations from individual singers' recordings - each set consisting of either low or high register singing. For each set, 3 one-second audio clips per vocal technique were randomly extracted from the singers vocalisations, provided they obeyed the following rules. The hierarchy of audio sampling for each set is presented in Appendix \ref{apdx:Stimuli}.

% pitch-matching and fry issues
Only one extract from a specific VocalSet recording could be used per set. Each low/high register set was assigned a `central pitch' which was determined by calculating a mean pitch value from all of the recordings for a given singer and lowering/raising this value by one standard deviation. The average pitch of each audio clip for a given set must be matched (within 2 semitones) to the assigned central pitch. If the average and central pitch do not match, a new audio clip is generated and the matched-pitch check is repeated. Often a singer's pitch for \textit{fry} utterances is a number of octaves below their other vocalisations (explained further in Section \ref{PotentialNoise}), making it impossible to match this technique's average pitch to the register's central pitch. If the central pitch cannot be matched after 20 audio clip generations, the low/un-pitched nature of the vocalisation is considered to be a feature of that singer's \textit{fry} technique, and the pitch-matching process is bypassed in these circumstances.

% adaptions to audio, 1 missing audio
There was a notably large variance in perceived volumes between singers and techniques. Extracted audio clips were therefore normalised to make the comparative task easier for participants. Due to an error in automated data collection, dissimilarities relating to 1 random audio clip were not saved correctly, leaving dissimilarity ratings for 14 audio files (instead of 15) to be used.

\subsection{Procedure}

% 60 participants, WAET, questionnaire, instruction, practise rounds, experiment, feedback, analysis
60 participants were recruited from audio/music-based academic departments as well as the author's music network, covering a wide range of music-enthusiasts. % demographic distributions for which are presented in the appendix. 
The study was conducted online using WAET.\footnote{Web Audio Evaluation Tool \citep{jillings2015}} Participants first completed a questionnaire\footnote{Includes `Perceptual ability' questions from GOLD-MSI \citep{mullensiefen2014}} which provided the demographic distributions presented in Appendix \ref{apdx:demoDist}. Participants were then instructed to listen to pairs of vocalisations and rate the dissimilarities between them on a continuous scale of 0 - 1 (see Appendix \ref{apdx:interface} and \ref{apdx:texts} for more details on the interface, instructive text and survey questions used). Participants were told their ratings should be irrespective of deviation in pitch (notes) or utterances (vowels). They were randomly assigned to listen to any of the 12 prepared vocal sets and were subjected to several practice rounds, allowing them to become familiar with the required task and diversity of timbres. Following this was the recorded experiment of 120 pairs of vocal recordings, plus 16 \textit{repeated} pairs for calculating intra-participant consistency. Participants were also invited to give open feedback at the end of the experiment regarding their experience and the techniques they used for rating. The dissimilarity ratings were then subjected to MDS, clustering, and statistical analysis.

\section{Results and Discussion} \label{secResDisc}

%Intro, voice ID, questionnaire and feedback, list of profile variables, intra-participant consistency
In this section we report statistically significant results from the data analysis. Participant data was divided into condition groups of singers, genders and registers. We refer to singers by their VocalSet ID, shortened to a `letter-index' format. Participants' questionnaire responses and feedback provided us with estimates for their perceptual ability (MSI scores); ranked scores for instrumental ability to reflect familiarity with music and singing (non-musician=0, musician=1 and singers=2); and task comprehension. We also calculated intra-participant consistency, generated by the repeated-rating comparisons using RMSE metrics.

\subsection{Clustering}
% missing data, correlation matrix for participant data, no outliers
Values for the missing data mentioned in Section \ref{dataSampling} were filled in with participant average ratings to create a uniform structure for all participant dissimilarity matrices. A correlation matrix was generated to show correlation coefficients between these matrices, upon which hierarchical clustering (HC) was performed. We observed that the data did not break off into minor clusters outside majority clusters for values of \textit{k}=1-5, indicating that outlier matrices of unusual behaviour such as inverse or binary raters were not detected.

%the hunt for ideal k value
To determine how participants perceptually clustered vocalisations, unsupervised clustering with HC algorithms was performed on dissimilarity matrices for \textit{k}=1-15. We generated optimal \textit{k} values using the elbow and silhouette score methods \citep{tan2018}. A lack of elbows in sum of squared error (SSE) plots implied that participants provided noisy dissimilarity data, or more likely, that distances between vocalisation techniques were fairly similar. Silhouette scores suggested the \textit{minimum} value of \textit{k}=2 implying that there is little salience among these singing techniques that would allow them to be segregated into more than 2 clusters.

Ground-truth labels were compared with HC predictions to determine an accuracy score.\footnote{Computed using scikit-learn's \texttt{adjusted\_rand\_score()} function} In total we use SSE, accuracy, and silhouette metrics to measure the performance of clustering. A Mann-Whitney test showed significant differences between conditions for cluster performance metrics for \textit{k}=5 (ground truth solution) and \textit{k}=2 (HC solution), as seen in Table \ref{c2c5diff}.

\begin{table}
\centering
\caption{
Mann-Whitney results table (\textit{p}$<$0.02). Each singer condition is accompanied with its mean value for the given metric. All 4 singer conditions in row 1, column 3 had significantly higher means compared to the singer condition in row 1, column 4 (and the opposite case for the fourth row). The \textit{U}-value reflects the effect size for each difference between conditions, proportional to the condition samples sizes (\textit{n}=10). \textit{k} indicates which number of clusters the metrics were calculated for.
}
\scalebox{0.76}{
\begin{tabular}{|l|l|l|l|l|}
\hline
\multicolumn{1}{|c|}{\textbf{k}} & \multicolumn{1}{c|}{\textbf{Metric}} & \multicolumn{1}{c|}{\textbf{Conditions with Higher Means}} & \multicolumn{1}{c|}{\textbf{Conditions with Lower Means}} & \multicolumn{1}{c|}{\textbf{U-value}} \\ \hline
2                                & Acc                               & M1=0.16,  M4=0.23,  F2=0.16,  F5=0.22                      & M2=0.02                                                   & 9.5, 3.0, 5.0, 6.0                    \\ \hline
2                                & Acc                               & F5=0.22                                                    & M1=0.16                                                   & 13.0                                  \\ \hline
2                                & Acc                               & M4=0.23, F5=0.22                                           & F3=0.09                                                   & 14.0, 8.0                             \\ \hline
5                                & Acc                               & M1=0.61                                                    & M2=0.09, M4=0.35, F3=0.20, F5=0.32                        & 5.0, 15.0, 10.5, 12.5                 \\ \hline
5                                & Acc                               & F5=0.32                                                    & M2=0.09                                                   & 11.5                                  \\ \hline
5                                & Sil                             & M1=0.36, M4=0.35                                           & M2=0.22                                                   & 5.0, 13.0                             \\ \hline
\end{tabular}
}
\label{c2c5diff}
\end{table}

This table shows compared conditions that exhibited significantly different distributions for a given metric (\textit{p}$<$0.02). We observe for \textit{k}=2, that compared to the majority of singers, M2's cluster accuracy is lower and F5's cluster accuracy is higher. One similarity between both \textit{k} solutions is that M2 scores lowest in accuracy, suggesting that this singer's singing techniques are particularly difficult to perceptually differentiate. For \textit{k}=5, M1 and M4 silhouette score distributions were higher than those of M2, implying that the clusters perceived for M1 and M4 vocalisations were better separated. There were no significant difference reported for SSE, and none for any metric between gender or register condition groups.

% participantProfile/ClusterMetrics correlation
We also tested for correlations across participant profile data and clustering quality metrics for \textit{k}=5 to see how participant profiles related to perceptions of the ground truth classes. Strong correlations existed between accuracy and silhouette scores (\textit{r}=0.60, \textit{p}$<$0.001), moderately negative between silhouette and SSE scores (\textit{r}=-0.51, \textit{p}$<$0.001) and faintly negative between accuracy and SSE (\textit{r}=-0.28, \textit{p}$<$0.05) due to the similarity in what is being measured.
Instrumental rankings had a weak correlation with MSI scores (\textit{r}\textsubscript{s}=0.30, \textit{p}$<$0.02) and moderate correlations with task comprehension (\textit{r}\textsubscript{s}=0.40, \textit{p}$<$0.01) and accuracy/silhouette (\textit{r}\textsubscript{s}=0.45/0.30, \textit{p}$<$0.001/0.02), which suggests that participants' level of instrumental familiarity allowed them to possess a more structured perception of the voice that was similar to the ground truth. The MSI scores were weakly correlated with SSE/Silhouette scores (\textit{r}=-0.30/0.27, \textit{p}$<$0.02/0.05), suggesting self-reported perceptual abilities were only vaguely reflected in cluster performance metrics.
SSEs were moderately correlated with intra-participant consistency (\textit{r}\textsubscript{s}=0.57, \textit{p}$<$0.001) showing that participants' inability to reproduce their ratings is indicative of loose clustering and unstable perceptual structures. Task comprehension was moderately correlated with accuracy (\textit{r}\textsubscript{s}=0.53, \textit{p}$<$0.001) and slightly negatively with SSEs (\textit{r}\textsubscript{s}=-0.34, \textit{p}$<$0.02), showing that task comprehension is indicative of good clustering metrics.

\subsection{Class Distance}

\begin{figure}
        \centering
        \includegraphics[width=0.4\textwidth]{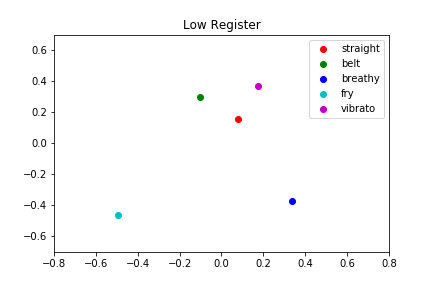}
        \vspace{0.1in}
        \includegraphics[width=0.4\textwidth]{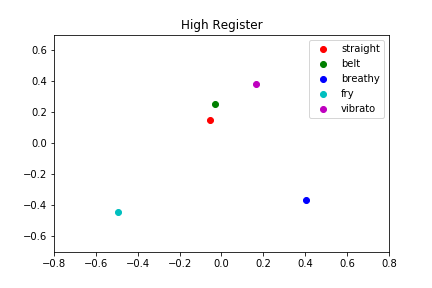}
        \hspace{0.01in}       
        \includegraphics[width=0.4\textwidth]{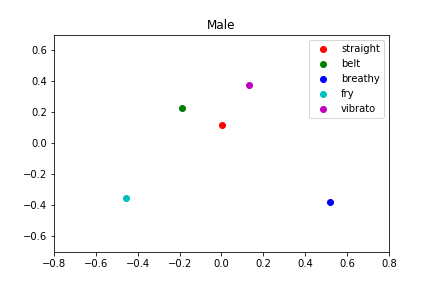}
        \vspace{0.025in}
        \includegraphics[width=0.4\textwidth]{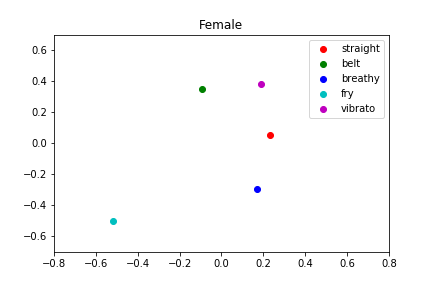}
        \caption{2D MDS plots representing the dissimilarities between the 5 ground truth labels for low register (top left), high register (top right), male (bottom left) and female (bottom right) conditions.
        }
        \label{mdsGraphsRegister}
\end{figure}

We averaged all dissimilarity ratings of the same class-pairs within each dissimilarity matrix (class-pair names will be abbreviated to 3 letters in this section). This allowed us to consider how class distances increase/decrease in the timbre space under different conditions.
Table \ref{classDistanceStats} shows statistically significant differences between conditions. There were many significant results for singer conditions, but these observations are not particularly meaningful without additional information and so are not included in the table (this is discussed further in Section \ref{concludeChap}).
Dissimilarities for class-pairs (Str - Bel), (Bel - Vib) and (Fry - Vib) were larger for low registers, which is mildly reflected in the corresponding MDS plot in Figure \ref{mdsGraphsRegister}. These plots should be noted with caution, as there are a considerable amount of high intra-class dissimilarities present in the data, 
which can raise issues when attempting to interpret MDS plots as they assume all intra-class dissimilarities to be zero. Certain class-pairs like (Bel - Bel) for males are an example of high intra-class dissimilarity values, implying that males' reproductions of similar vocal techniques are perceptually diverse for \textit{belt}. This is also the case for females' \textit{breathy} and \textit{vibrato} classes.
Both Table \ref{classDistanceStats} and Figure \ref{mdsGraphsRegister} reflect larger (Str - Bre) distances for males and larger (Fry - Vib) distances for females.
Lastly, Figure \ref{mdsGraphsRegister} shows that \textit{belt}, \textit{vibrato} and \textit{straight} techniques are perceptually similar across both register and gender conditions, while \textit{fry} is consistently most dissimilar from the other classes.

\begin{table}
\centering
\caption{Mann-Whitney test results comparing class distances between conditions (similar layout to Table \ref{c2c5diff}). Class names are abbreviated to 3 letters.}
\scalebox{0.75}{
\begin{tabular}{|l|l|l|l|l|}
\hline
\textbf{Condition Group} & \textbf{Class Pair} & \textbf{Condition with Higher Means} & \textbf{Condition with Lower Means} & \textbf{U-value} \\ \hline
Register                 & Str - Bel           & low=0.59                              & high=0.45                            & 263.0            \\ \hline
Register                 & Bel - Vib           & low=0.58                              & high = 0.46                          & 301.0            \\ \hline
Register                 & Fry - Vib           & low = 0.81                            & high = 0.70                          & 304.0            \\ \hline
Gender                   & Bel - Bel           & male=0.32                             & female=0.24                          & 303.5            \\ \hline
Gender                   & Str - Bre           & male=0.67                             & female=0.51                          & 235.0            \\ \hline
Gender                   & Fry - Vib           & female=0.82                           & male=0.70                            & 275.0            \\ \hline
Gender                   & Bre - Bre           & female=0.21                           & male=0.16                            & 291.0            \\ \hline
Gender                   & Vib - Vib           & female=0.22                           & male=0.16                            & 298.5            \\ \hline
\end{tabular}
}
\label{classDistanceStats}
\end{table}

\subsection{Potential Sources of Noise} \label{PotentialNoise}

VocalSet has some shortcomings that may contribute towards noise in the data. Many recordings contain multiple techniques, despite being labelled exclusively as one. The quality of performances seems to vary considerably between singers. Due to the nature of the \textit{fry} technique and variance in performance style, its pitch is often several octaves below the singer's intended pitch (and the dataset's implied pitch label). The perceived volume also differs between recordings.

% participant feedback implies noise
Participants reported factors that influenced or dictated their dissimilarity evaluations which are summarised as: performers lack of control, soft/harshness, clean/dirtiness, distortion, dynamics, temporal pitch variation, subglottal pressure, larynx placement, resonance, total amount of notes per sample, melody, emotion, register mechanics and assumed class types. Many of these imply that there is a considerable degree of uncertainty regarding the dissimilarity evaluation task. It is reasonable to believe that these issues may have also caused a significant level of noise in the results.

% maybe mention feedback from pilot too?
\section{Conclusion} \label{concludeChap}

% participant profile results
In this study we have shown that participants' instrumental ability and consistency in their ratings supported similarity between their perceptions of the voice and the ground truth labels, as well as cluster cohesion/separation.
We have also shown that there are subtle similarities and differences in the timbre space between genders and registers and that intra-class variance for female vocalisations are wider than for males.
Clustering analysis however, showed that participants' data did not separate into clusters easily.
Participant feedback analysis suggested that instructions given to participants could be revised to better articulate how dissimilarity ratings should be evaluated. Part of our assumption was that very minor deviations in pitch would have a negligible affect on timbre perception. However, as pitch variance was distracting for participants, it may be worth focusing soley on sustained single pitches in future work.

In this study, there were significant differences in clustering performances and class distances between singers. Reasons for these are best explored with joint analysis of acoustic descriptors and dissimilarities for vocalisations, allowing us to understand how singers' acoustic attributes influence clustering behaviour, while also assisting in the interpretation of the MDS dimensions. In future work, we also intend to use the MDS-generated timbre maps for regularisation in generative neural networks for inferring a model of vocal timbre in accord with human perception.

\bibliographystyle{apacite}
\bibliography{myBib20200821c.bib}

\appendix

\section{Hierarchical Structure of Session Sets} \label{apdx:Stimuli}

\begin{figure}[H]
    \centering
    \includegraphics[width=0.9\textwidth, height=6cm]{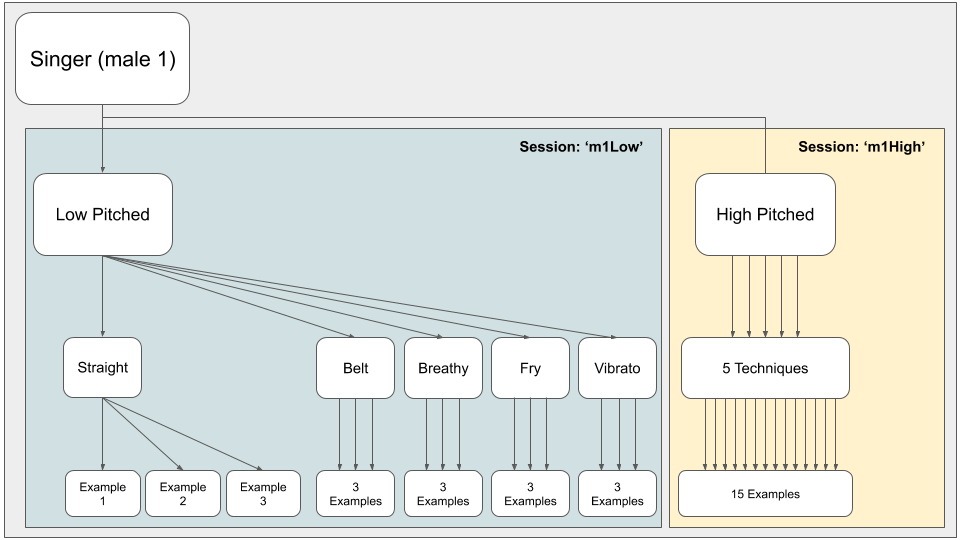}
    \caption{Breakdown of the sampling structure (per singer) used to generate stimuli for perceptual evaluations.}
    \label{samplingScheme}
\end{figure}

% Instead of using \pagebreak, \newpage shouldn't push the figure to the bottom (i.e. vertical pad everything on this page)
% \newpage

\section{Participant Demographic Distributions} \label{apdx:demoDist}

\begin{figure}[H]
    \centering
    \includegraphics[width=0.8\textwidth]{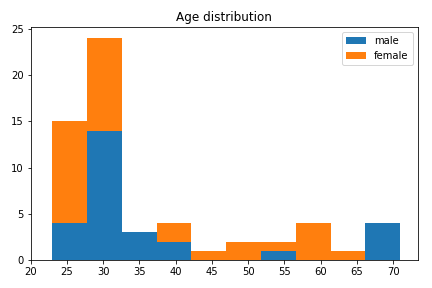}
    \caption{Distribution of participant age and genders. Participant age $\mu = 36.45, \sigma = 13.69$}.
    \label{samplingScheme}
\end{figure}

\begin{figure}[H]
    \centering
    \includegraphics[width=0.8\textwidth]{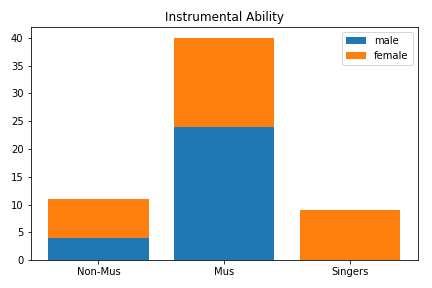}
    \caption{Number of non-musicians (non-mus), musicians (mus) and musicians with singing as their primary instrument (singers). No male participants considered the voice as their primary instrument.}
    \label{samplingScheme}
\end{figure}

\section{Interface View} \label{apdx:interface}

\begin{figure}[H]
    \centering
    \includegraphics[width=0.8\textwidth]{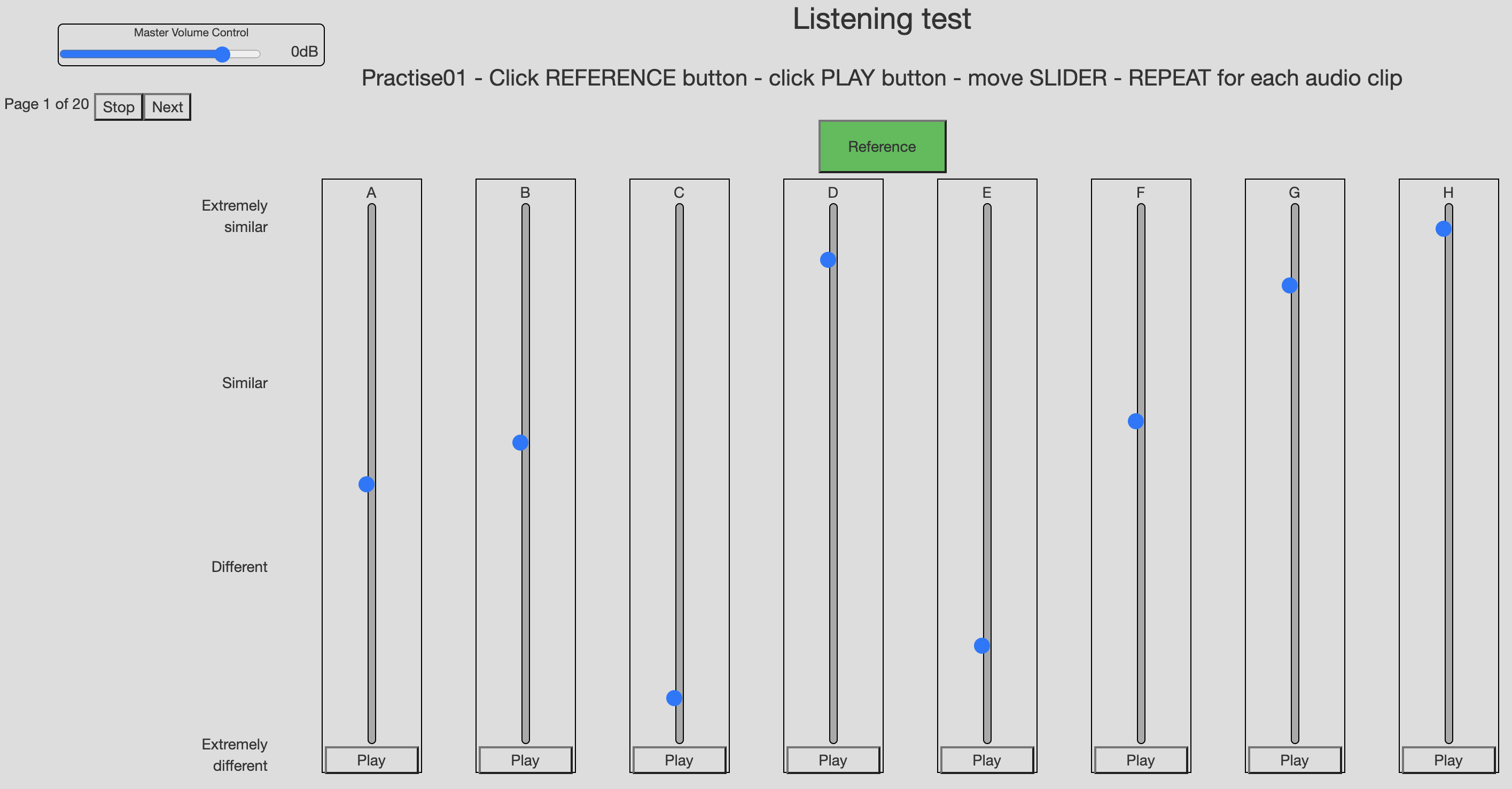}
    \caption{View of interface used by participants for rating dissimilarities between reference (remains the same for each page) and comparative audio clips.}
    \label{samplingScheme}
\end{figure}

\section{Study Texts} \label{apdx:texts}

\subsection{GOLD-MSI Perceptual Ability Questions}

The following questions were extracted from the GOLD-MSI `Perceptual Abilities' subset \citep{mullensiefen2014}, and uses a 7-point agreement scale:

\begin{enumerate}
    \item I am able to judge whether someone is a good singer or not.
    \item I usually know when I'm hearing a song for the first time.
    \item I find it difficult to spot mistakes in a performance of a song even if I know the tune.
    \item I can compare and discuss differences between two performances or versions of the same piece of music.
    \item I have trouble recognizing a familiar song when played in a different way or by a different performer.
    \item I can tell when people sing or play out of time with the beat.
    \item I can tell when people sing or play out of tune.
    \item When I sing, I have no idea whether I'm in tune or not.
    \item When I hear a music I can usually identify its genre.
\end{enumerate}

\subsection{Additional Questions}

Apart from question 6, the following questions were taken before the study was conducted:

\begin{enumerate}
    \item Please indicate what listening equipment you intend to use for this experiment (Headphones are preferrable). If you wish to change your setup, please do so before continuing and refresh this page [Inbuilt speakers, external speaker, ear/headphones]
    \item How would you assess your current listening environment on a scale of 1 (very noisy) to 5 (very quiet)? [Integer]
    \item Please provide your age in the space below. [Integer]
    \item Please provide your gender identity in the space below. [Male, female]
    \item What instrument are you best at playing? [Instrument name]
    \item Do you have any other comments regarding your evaluations, or any other aspect of the study? [Post-study question, open-ended response]
\end{enumerate}

\subsection{Task Description}

The following text quotes the instructions given to participants regarding the task required of them:

\begin{quote}{\small}
We are interested in measuring how differently listeners perceive the sounds of a singer's voice when undergoing various singing techniques. In this experiment you will be comparing between multiple, unedited and unprocessed recordings of one individual singer. Your task is to rate how similar or different the singer's sustained vocalisations sound to you, due to different singing techniques. The challenge therefore, is to rate vocal similarities IRRESPECTIVE of the singer's changes in pitch (notes) and utterance (vowels) between recordings.
\end{quote}

\end{document}